\documentclass[10pt,letterpaper]{article}
\usepackage{opex3}

\begin{document}

\title{A vibration-insensitive optical \\cavity and absolute determination of its ultrahigh stability}

\author{Y. N. Zhao, J. Zhang, A. Stejskal, T. Liu, V. Elman,\\ Z. H. Lu, and L. J. Wang$^\ast$}

\address{Max Planck Institute for the Science of Light, \\and Institute of Optics,
Information and Photonics, University Erlangen-Nuremberg,
G\"unther-Scharowsky-Str. 1 / Building 24,\\ 91058 Erlangen, Germany}

\email{Lijun.Wang@mpl.mpg.de} 



\begin{abstract}
We use the three-cornered-hat method to evaluate the absolute frequency stabilities of three different ultrastable reference cavities, one of which has a vibration-insensitive design that does not even require vibration isolation. An Nd:YAG laser and a diode laser are implemented as light sources. We observe $\sim1$ Hz beat note linewidths between all three cavities. The measurement demonstrates that the vibration-insensitive cavity has a good frequency stability over the entire measurement time from $100$ $\mu$s to $200$ s. An absolute, correlation-removed Allan deviation of $1.4\times10^{-15}$ at $1$ s of this cavity is obtained, giving a frequency uncertainty of only $0.44$ Hz.  
\end{abstract}



\section{Introduction}
Ultra-narrow linewidth lasers play important roles in high precision metrology, such as high-resolution spectroscopy, optical frequency standards, and fundamental tests of physics \cite{science2008_Al, science2008_Sr, prl2005_Yb, lp2007_Wang}. To achieve such a frequency stabilized local oscillator, the laser is typically servo locked to a vibration isolated, high-finesse cavity by using the Pound-Drever-Hall (PDH) technique \cite{apb1983_PDH}. Since the laser frequency follows the modal frequency of the reference cavity, the stability of the reference cavity itself becomes essential to achieve ultra-high frequency stability \cite{josab5_Hall}. Environmental disturbance is often one of the dominant noise sources. Seismic vibrations can be transmitted to the cavity spacer via its structural support. Such vibration leads to forces which deform the cavity and change its length. Length fluctuation then acts as a broadband noise source that modulates the laser frequency, resulting in linewidth broadening. Some interesting works have already shown that the sensitivity of the cavity to the environmental mechanical disturbance can be minimized by optimizing cavity geometry and mounting method \cite{prl_Young, ol2007_Ye, pra2008_Gill}. 

To verify the achieved laser stability, a standard method is to heterodyne beat the stabilized laser with another similar laser, and take $1/\sqrt{2}$ of the combined beat frequency stability as the individual laser stability \cite{ol2007_Ye, pra2008_Gill, ol2004_Gill, pra2008_Hansch}. However, this method is invalid if the two lasers have dissimilar performance levels. To solve this problem, we have carried out an experiment to directly characterize the absolute frequency stability of three different reference cavities by simultaneously cross-beating three laser beams that are frequency stabilized to these three reference cavities, following the idea of the ``three-cornered-hat'' method \cite{ol2009_Liu}. In the reported work, only one Nd:YAG laser was used to generate three different laser beams that were independently locked to three reference cavities. As a consequence, there might still exist an ambiguity in separating common mode noise from individual cavity noise. 

In this paper, we report a much improved experiment that resolves these ambiguities. The improvement is two-fold: the use of two primary light sources removes common phase noise; and a vibration insensitive cavity design is implemented that does not even require vibration isolation under normal laboratory seismic conditions. To improve vibration insensitivity of a horizontally placed cavity, special mounting points (Airy points) are found and a vertical acceleration sensitivity of less than $10^{-14}$ s$^2$ is achieved. To characterize the frequency stability of this cavity, we use two other different design ultrastable cavities (cavity $1$ and cavity $2$), together with the newly designed cavity (cavity $3$), to perform a three-cornered-hat measurement. All three cavities are coated for a wavelength of $946$ nm, whose fourth harmonic at $236.5$ nm is used for ultrahigh resolution spectroscopy of the $5s^2$ $^1$S$_0$ -- $5s5p$ $^3$P$_0$ transition of a trapped $^{115}$In$^+$ ion. Cavity 1 and cavity 2 are put on top of two independent active vibration isolation (AVI) stages, while cavity 3 is only placed on a regular breadboard. In this work, instead of using only one laser source, we use two independent laser sources to suppress the effect of common noise. Two light beams derived from an Nd:YAG laser are locked to cavity $1$ and cavity $2$, respectively. A diode laser source, placed in another laboratory, is pre-stabilized to a reference cavity, transferred through a fiber link, and then locked to cavity $3$. By cross beating the three stabilized laser beams simultaneously, the absolute frequency stabilities of the individual cavities are measured. An absolute frequency stability of $1.3\times10^{-15}$ at $0.4$ s is observed for cavity $3$, even without vibration control. In classical three-cornered-hat method the three cavities are assumed to be uncorrelated. This hypothesis might not be correct, and can cause the appearance of negative frequency stability in certain cases. In this work, we solve this problem by calculating the covariance matrix of the three beat frequencies. The results show that correlations do exist between cavities at the time scale from $10$ s to $200$ s, mainly due to the same experimental surroundings of the three cavities. 

Since the vibration-insensitive cavity design is central to this paper, a detailed discussion of the cavity design is presented in section $2$. In addition to the information about the three cavities, this section also describes the frequency locking schemes, and the fiber noise compensation setup. Section $3$ presents the experimental results of the three-cornered-hat measurements, including the short term linewidth measurements, and the Allan deviations of the cross-beating among the three reference cavities, from which we can deduce the absolute Allan deviations of the individual cavities. We also investigate the correlations of the three cavities, and present the results of absolute, correlation-removed Allan deviations of the individual cavities. The results are summarized in section $4$.

\section{Experimental Setup}
\subsection{Reference cavities}
\begin{figure}[htb]
\centering\includegraphics[width=9cm]{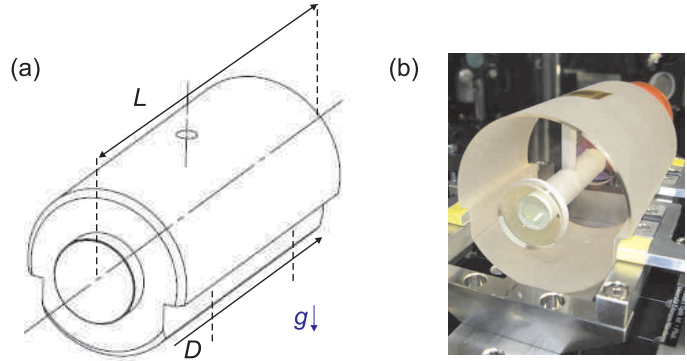}
\caption{(a) Three-dimensional view of the cavity $3$ geometry. $L$ is the cavity length, excluding the two mirrors, position $D$ is along the cavity axis. (b) Picture of Cavity $3$.}
\label{CavDesign}
\end{figure}

An optical reference cavity serves as the most essential element of an ultra-narrow linewidth laser system. As a consequence, it needs to be isolated from environmental disturbance, such as acoustic noise and mechanical (seismic) vibration. This can be done either by putting the cavity on a vibration isolation stage, or by designing a vibration insensitive cavity, or both. In the present case, cavity 1 and cavity 2 are cylindrical design. They are more sensitive to vibration perturbations, and have to be put on top of vibration isolation stages. Further information about this two cavities can be found in Ref. \cite{oc2002_Nevsky, apb2007_Liu}. Cavity 3 is a new design with intrinsic vibration immunity, similar to that of Ref. \cite{apb2006_Nazarova}. The geometry of the cavity is shown in Fig. \ref{CavDesign}(a), and a picture of the cavity is shown in Fig. \ref{CavDesign}(b), where $L$ is the cavity length excluding the two mirrors, position $D$ is along the cavity axis. Square ``cutouts'' are made on the bottom of the cylindrical spacer and the cavity is supported at four points. The cutouts compensate for vertical forces, and vibration insensitivity is achieved through the special cavity shape and symmetrical mounting at the Airy points.

With the help of a commercially available software (Comsol Multiphysics), we perform a detailed numerical analysis of the reference cavity configuration by finite-element analysis to optimize its design. The cavity spacer and the optically bonded mirror substrates are made of ultra-low-expansion glass (ULE), with density of $2.21\times10^3$ kg/m$^3$, elastic modulus of $6.67\times10^{10}$ N/m$^2$, and Poisson's ratio of $0.17$. The spacer has a length of approximately $10$ cm and an outside diameter of approximately $10$ cm. Since the vibration frequencies that significantly contribute to the cavity instability are less than $10$ Hz, they can be considered as dc relative to the first structural resonance of the cavity, which is about $10$ kHz. Therefore, we choose a static stress-strain model with a gravity-like force applied on the cavity in the vertical direction, as shown in Fig. \ref{CavDesign}(a). To reduce modeling computation time, we take advantage of the symmetry of the cavity and only calculate a quarter section of the cavity excluding the mount. As only symmetric solutions are considered, the planes of symmetry are constrained to move only along the plane surfaces. The support is modeled as a point constrain on the horizontal cutout surface in order to prevent the cavity spacer from translational or rotational movements.  

\begin{figure}[htb]
\centering\includegraphics[width=7cm]{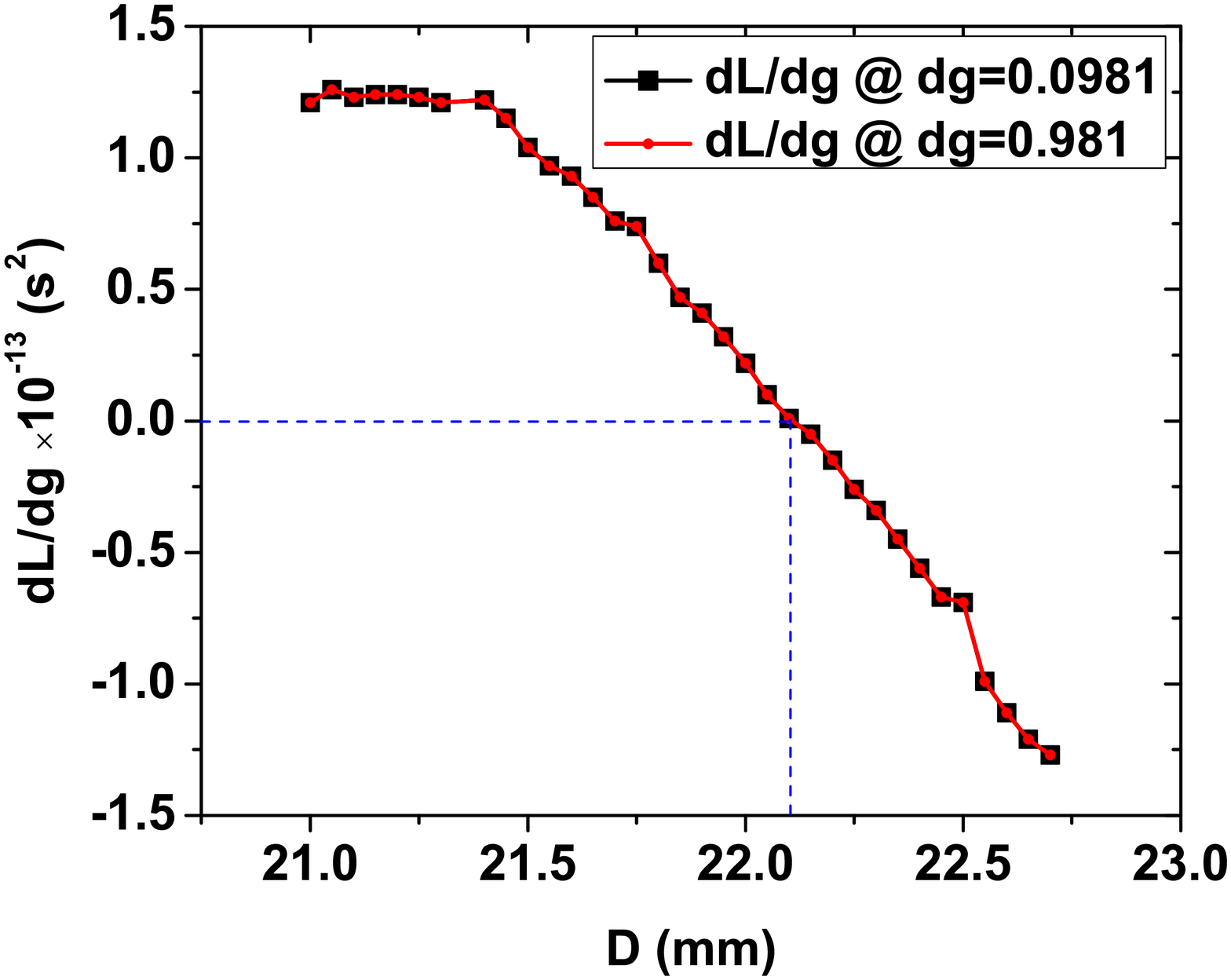}
\caption{Typical result of d$L/$d$g$ at different supporting point of $D$, where d$L$ presents the change in cavity length, d$g$ stands for a small change in the acceleration $g$. The black square curve and the red circle curve indicate the values of d$L$/d$g$ with the different displacements at $D$ when d$g=0.0981$ m$/$s$^2$, and d$g=0.981$ m$/$s$^2$, respectively.}
\label{dLdg}
\end{figure}

The simulated results of the cavity acceleration sensitivity d$L/$d$g$ at different supporting points $D$ is shown in Fig. \ref{dLdg}. Here d$L$ presents the change in cavity length caused by d$g$, a small change in the acceleration $g$. Both d$g=0.0981$ m$/$s$^2$ and d$g=0.981$ m$/$s$^2$ are calculated, and the results are nearly the same, showing good linearity. The results indicate that there exists an optimal supporting position where two cavity mirrors are parallel and insensitive to the vertical accelerations. At this point, the cavity deforms under the gravity, but the axial distance between the centers of the mirror inner surfaces remains unchanged. Furthermore, the axial displacement can be calculated as a function of support position and cut depth. We choose a cut depth of $5$ mm at a vertical position of $4.5$ mm below the mid-plane. The calculated result shows a d$L/$d$g<10^{-14}\ \mathrm{s}^2$ acceleration sensitivity.

We place cavity $3$ on a home-made stainless steel base with four supporting points. The whole assembly sits inside a stainless steel vacuum chamber at a pressure of $10^{-8}$ mbar. The vacuum chamber is placed inside a temperature-stabilized aluminum box with a digital PID control. The temperature is stabilized at $27$ $^{\circ}$C with a root-mean-square (RMS) fluctuation of less than $1$ mK. To reduce the influence of acoustic noise, the vacuum chamber is surrounded by plastic foams containing a layer of lead septum. The cavity and other optical components sit on top of a 10-cm-thick breadboard without any vibration control. The three platforms for cavities 1, 2 and 3 are located on top of an unfloated optical table inside an acoustically isolated cabin (lab $1$).

\begin{figure}[htb]
\centering\includegraphics[width=7cm]{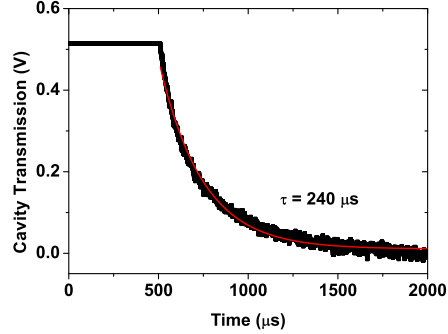}
\caption{The cavity ring down signal of cavity $3$, shown with the black curve. The red curve is the exponential fitting of the decay time.}
\label{CavRingDown}
\end{figure} 

To calculate the finesse of cavity 3, we measure the cavity ring down (CRD) time of the cavity. The result is shown in Fig. \ref{CavRingDown}. Similar measurements are performed for cavity 1 and cavity 2. From the measured cavity free spectrum range (FSR) and CRD time, we can calculate the corresponding cavity length, linewidth, and finesse. The results are summarized in Table \ref{parameters}.

\begin{table}[htbp]
\renewcommand\arraystretch{1.5}
\begin{center}
\caption{Summary of the cavities specifications.}
\label{parameters}
\begin{tabular}{|c|cccccc|}\hline
Cavity No. &Material&FSR      &Length   &CRD         &Cavity Linewidth &Finesse\\ \hline
$1$        &Zerodur &$559$ MHz&$26.8$ cm&$24$ $\mu$s &$8$ kHz          &$7\times10^{4}$ \\
$2$        &ULE     &$750$ MHz&$20.0$ cm&$27$ $\mu$s &$6.6$ kHz        &$1.2\times10^{5}$ \\
$3$        &ULE     &$1.6$ GHz&$9.4$  cm&$240$ $\mu$s&$660$ Hz         &$2.4\times10^{6}$ \\\hline
\end{tabular}
\end{center}
\end{table}

\subsection{Laser locking}
\label{Locking}
\begin{figure}[htb]
\centering\includegraphics[width=13cm]{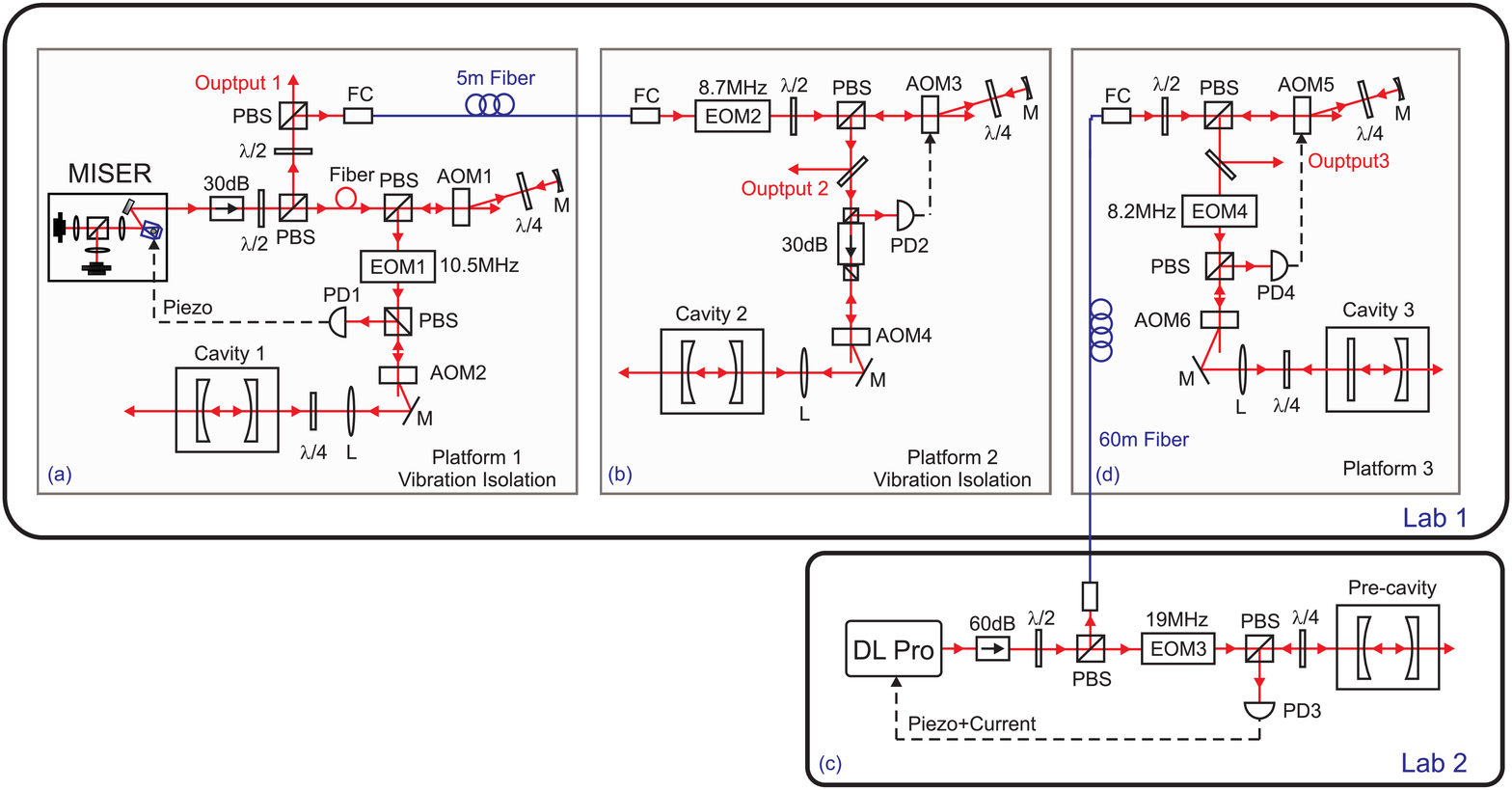}
\caption{(a) Schematic of the MISER locking to cavity $1$. (b) Schematic of the MISER locking to cavity $2$. (c) Schematic of the diode laser locking to a pre-cavity in lab $2$. (d) Schematic of the diode laser locking to cavity $3$. EOM, electro-optic modulator; AOM, acousto-optic modulator; $\lambda/2$, half waveplate; $\lambda/4$, quarter waveplate; PBS, polarization beam splitter; PD, photodiode; FC, fiber coupler; M, mirror; L, lens.}
\label{Lockingsetup}
\end{figure}

The light sources we used are a monolithic isolated end-pumped ring Nd:YAG laser (MISER) (Innolight Mephisto QTL $500$ NE) \cite{lp2007_Wang, apb2007_Liu}, and an extended cavity diode laser (Toptica Photonics DL Pro), both operating at $946$ nm. The MISER has an intrinsically high frequency stability and low amplitude noise. It has a high output power of $300$ mW. A total power of only $5$ mW from the laser is used to lock the laser frequency to cavity $1$ by the PDH locking technique, as shown in Fig. \ref{Lockingsetup}(a). The main output of the stabilized laser is used as a local oscillator for an optical frequency standard based on a single trapped indium ion. A 20-cm single-mode fiber is used to clean up the weak laser beam's spatial mode. A $160$ MHz acousto-optical modulator (AOM1) is double-passed to shift the laser frequency to the desired TEM$_{00}$ cavity mode. The laser beam then passes through an electro-optical modulator (EOM1), which generates two $10.5$ MHz sidebands around the carrier frequency for PDH frequency locking. The temperature of EOM1 is stabilized to avoid thermal-induced fluctuations in the polarization of the laser light. Approximately $30$ $\mu$W power is directed into cavity $1$, with a proper confocal lens for cavity mode-matching to obtain a maximum coupling efficiency. The reflected light from cavity 1 is detected by a photodiode (PD$1$). An additional AOM (AOM$2$) is used as an optical isolator to prevent standing waves on PD$1$. The signal from PD$1$, after demodulation, is used as an error signal to feedback control the frequency of the MISER. This feedback signal is applied to a piezo that is glued on top of the MISER crystal. It deforms the crystal, and changes the optical path length of the oscillator. To increase the dynamic range of the feedback control, an additional thermal controller is servo controlled to change the temperature of the laser crystal. This combined feedback system typically can maintain continuous frequency locking for a full day. Part of the main beam (output $1$) is split off and reserved for the three-cornered-hat measurement.

$20$ mW of the remaining power of the MISER is transferred to the second platform via a 5-m-long single-mode fiber for cavity $2$. After passing through EOM2 to add sidebands at $8.7$ MHz, light from the fiber output double-passes through AOM3, as shown in Fig. \ref{Lockingsetup}(b). AOM3 is used to bridge the frequency difference between the two closest TEM$_{00}$ modes of cavity $1$ and $2$, and is also used for PDH locking to cavity 2. Double-passing arrangement allows fast frequency corrections without causing beam deviation. The light after double-passing AOM$3$ is separated into two parts. A small part of light, approximately $30$ $\mu$W, is mode-matched into cavity $2$. The servo bandwidth for AOM3 locking is $\sim15$ kHz. The other part of the light (output $2$) is used for the three-cornered-hat measurement.   

The extended cavity diode laser is placed in another lab (lab $2$). Before transmitting the diode laser light to lab $1$, we pre-stabilize the laser to an optical cavity with a finesse of $73\ 000$ by means of the PDH locking technique, as shown in Fig. \ref{Lockingsetup}(c). The linewidth of the laser diode is thus reduced from $\sim1$ MHz level to a few hundreds Hz with a locking bandwidth of $\sim2$ MHz. The pre-locking cavity is also made of low thermal expansion material, and a low vacuum environment ($2\times10^{-7}$ mbar) is maintained to decrease the cavity mode drift. The pre-stabilized light is then transferred through a 60-m-long single-mode fiber to lab $1$. To further lock the diode laser to cavity $3$, a similar locking scheme as used for cavity $2$ is arranged, as shown in Fig. \ref{Lockingsetup}(d). Approximately $2$ mW stabilized light (output $3$) is reserved for the three-cornered-hat measurement.

\subsection{Fiber noise compensation}
As indicated in Fig. \ref{Lockingsetup}, AVI systems are used for cavity $1$ and $2$, except for cavity $3$. By using an AVI system, the amplitude of the mechanical vibrations can be suppressed by more than $10$ dB, measured from DC to $100$ Hz frequency range. The three-cornered-hat measurement is performed on platform $3$ by cross beating the three laser beams, output $1$, $2$, and $3$, that are frequency stabilized to the respective reference cavities. Fast photodiodes (EOT ET-$3000$A) are used for beat signal detections.

\begin{figure}[htb]
\centering\includegraphics[width=7cm]{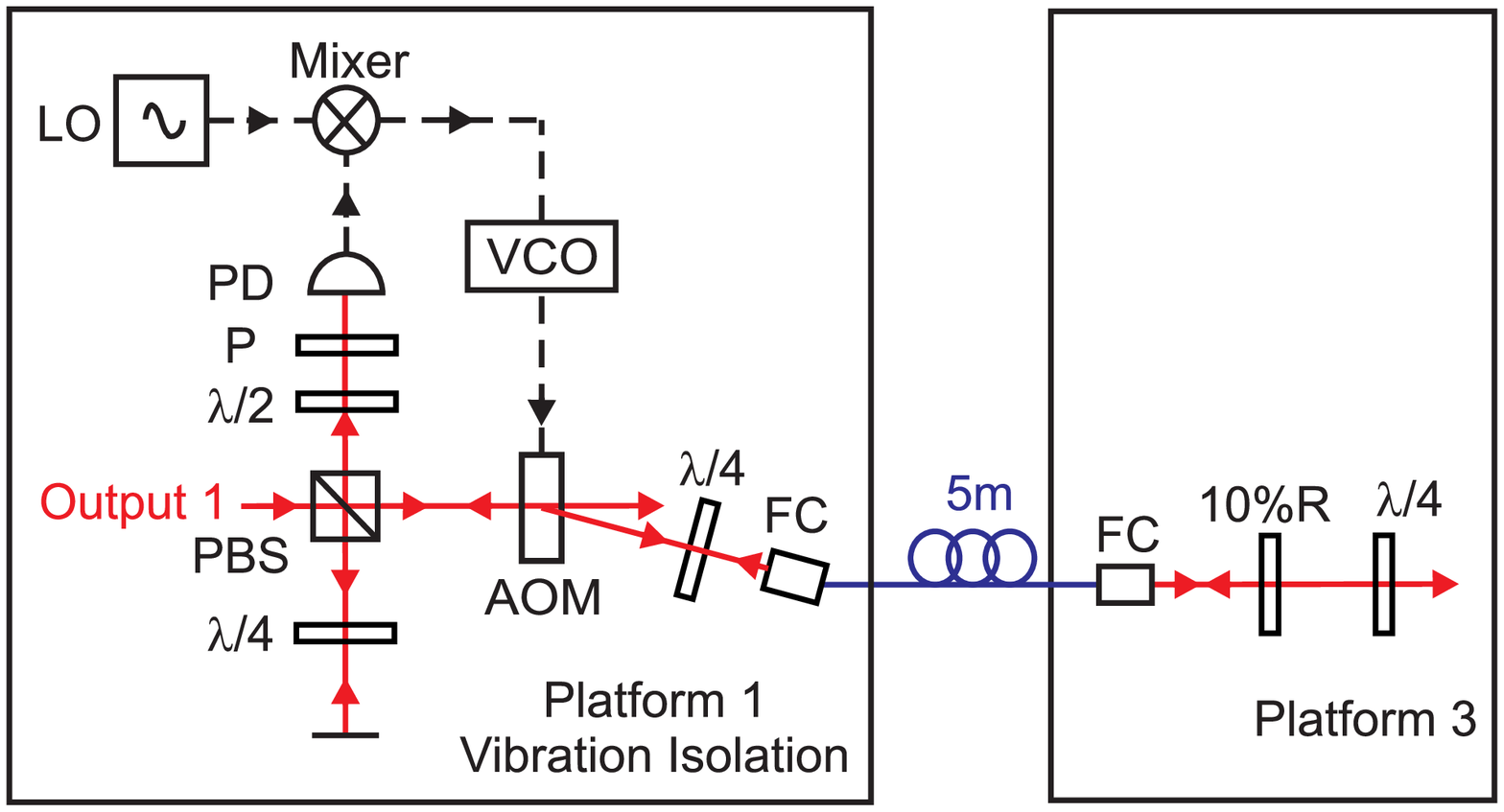}
\caption{Schematic of the fiber noise compensation. Output $1$, the laser beam stabilized to cavity $1$; AOM, acousto-optic modulator; $\lambda/2$, half waveplate; $\lambda/4$, quarter waveplate; PBS, polarization beam splitter; P, polarizer; PD, photodiode; FC, fiber coupler; $10$\%R, $10$\% reflection beam splitter.}
\label{FNC}
\end{figure}

Output $1$ and $2$ travel from their vibration isolated platforms via two 5 m long single-mode fibers to platform $3$. To retain the high spectral purity of the stabilized lights, the fiber noise caused by acoustic, mechanical, and thermal perturbations are canceled with a compact scheme that is similar to the one used in Ref. \cite{ls1999_FNC} to avoid any linewidth broadening. The fiber noise compensation setup for output $1$ is shown in Fig. \ref{FNC}, which is the same for output $2$. The stabilized light propagates through an AOM, and the first-order, frequency-shifted light goes through the fiber. $10$\% of this light is retroreflected back through the fiber and the AOM, and is heterodyne mixed with the original input light. A phase-lock loop is used to control the AOM driving frequency so that the beat signal is phase-coherent with a local reference frequency. The frequency of the light at the fiber remote end can be expressed as
\begin{equation}
f_{remote}=f_0+f_{AOM}+f_{noise}+f_{mod},
\label{remote}
\end{equation}
where $f_{remote}$ is the laser frequency at the fiber remote end, $f_0$ is the frequency of the original incoming light, $f_{AOM}$ is the center driving frequency of the AOM, $f_{noise}$ is the noise frequency picked up by the fiber, and $f_{mod}$ is the modulation frequency imposed on the AOM by the phase-lock loop. The frequency of the retroreflected light after double-passing the AOM and the fiber, $f_{local}$, is
\begin{equation}
f_{local}=f_0+2(f_{AOM}+f_{noise}+f_{mod}).
\label{local}
\end{equation}
Here we assume that light passing through the fiber picks up the same amount of phase noise in both directions and that the counterpropagating waves are transmitted independently of each other. The beat signal, $2(f_{AOM}+f_{noise}+f_{mod})$, is tightly phase-locked to a local reference frequency at $2f_{AOM}$, such that $f_{noise}+f_{mod}$ remains zero. According to Eq. (\ref{remote}) and (\ref{local}), it is clear that when the phase-lock loop is closed, both local and remote frequencies at the two ends of the fiber can retain their high spectral purity. The AOM is used locally, as shown in Fig. \ref{FNC}, so that the cables for the phase-lock loop can be as short as possible to avoid picking up extra noise. 

As mentioned in section \ref{Locking}, the pre-stabilized diode laser is transferred through a $60$ m fiber to lab 1. The fiber noise caused by this long fiber does not need to be compensated, because both the laser spectral noise and the fiber noise are suppressed by AOM5. As a consequence, output 3 does not pick up any fiber noise from this $60$ m long fiber. 

\section{Experimental results}
\subsection{Classical three-cornered-hat measurement}
\label{AllanDev}
\begin{figure}[htb]
\centering\includegraphics[width=13cm]{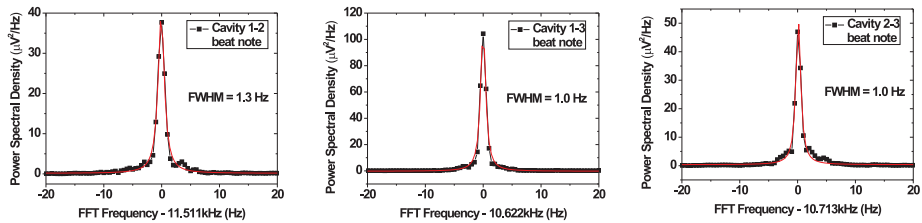}
\caption{Optical beat signals between the three cavities. For the SR785 FFT spectrum analyzer, a frequency resolution of $0.5$ Hz is used by choosing a $400$ Hz span with a $2$ s acquisition time. For each beat note, 10 scans are recorded and averaged. The red curves are Lorentzian fit results.}
\label{FFT}
\end{figure}

When the three laser beams are stabilized to the respective cavities and the fiber noise compensation loops actively running, three cross beat frequencies are obtained at approximately $88$ MHz, $304$ MHz, and $392$ MHz for cavity $1$-$2$, $1$-$3$, and $2$-$3$, respectively. We then down convert the beat frequencies to kHz level and use low pass filters to clean up the signals. To estimate the short-term linewidths of the three cavities, we use a fast Fourier transform (FFT) spectrum analyzer (Stanford Research Systems SR785) to analyze the three cross beat frequencies. A frequency resolution of $0.5$ Hz is achieved by choosing a $400$ Hz span with a $2$ s acquisition time. For each beat frequency, 10 scans are recorded and averaged, as shown in Fig. \ref{FFT}. The black curves are power spectral densities of the beat signals, and the red curves are Lorentzian fittings. The full width at half maximum (FWHM) of all the beat signals is at the $1$ Hz level. It can be seen that the beat frequency between cavity $1$-$2$ has the broadest linewidth, and the best beat note spectral shape is obtained between cavity $1$-$3$ with minimum spectral noise. Since the three beat signals are not taken at the same time, they only serve as indicators of these cavities' performance.

\begin{figure}[htb]
\centering\includegraphics[width=7cm]{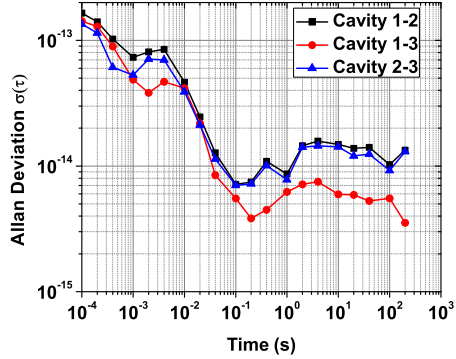}
\caption{The Allan deviations of the beat frequencies between three reference cavities. The linear frequency drifts of approximately $1$ Hz/s are removed during data processing.}
\label{ADBeat}
\end{figure}

To further analyze the stabilities of the three reference cavities, a three-cornered-hat measurement is performed. The three down-converted beat signals are recorded by three frequency counters (Agilent 53132A), for gate time longer than $1$ s, and by three fast analog-digital converters (Gage CompuScope 12400), for gate time shorter than $1$ s. As for the Gage CompuScope cards, to achieve a resolution of $100\ \mu$s, we choose a $100$ kHz sampling rate, and synchronize the three channels to one common external trigger. The Allan deviations of the three data sets are calculated, and are shown in Fig. \ref{ADBeat}. Linear frequency drifts of approximately $1$ Hz/s are removed during data processing. We clearly observe a much better beat frequency stability for cavity $1$-$3$ than those of cavity $1$-$2$ and $2$-$3$ over the entire time scale. The results agree with the linewidth measurement results. The last two beat frequency stabilities have more or less the same shape, mainly due to the relative poor performance of cavity $2$. The beat frequency of cavity $2$-$3$ is still slightly better than that of cavity $1$-$2$ at all time scales, which suggests that the performance of cavity 3 is better than that of cavity 1.
 
\begin{figure}[htb]
\centering\includegraphics[width=12cm]{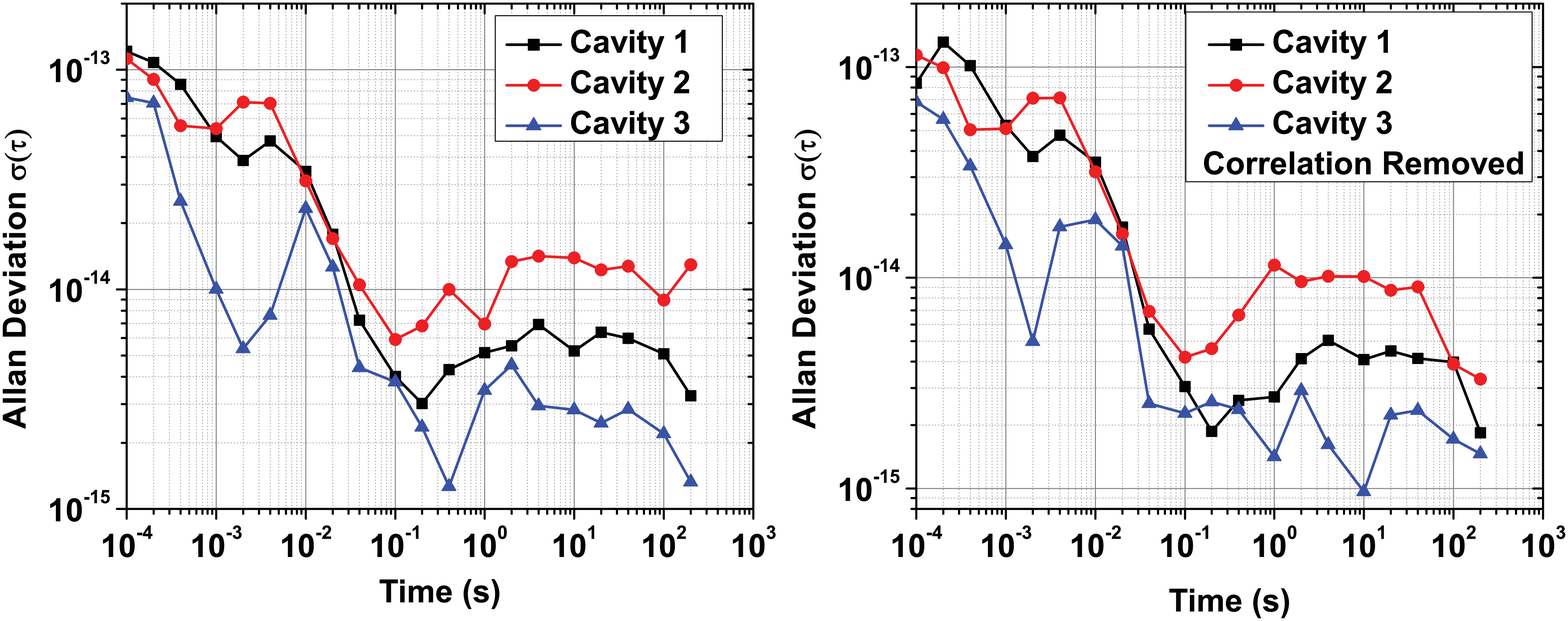}
\caption{(a) The Allan deviations of the individual reference cavities calculated from the classical three-cornered-hat measurement. (b) The Allan deviations of the individual reference cavities with the correlations removed according to Ref. \cite{ieeetim1993_Premoli}.}
\label{ADSingle}
\end{figure}

To characterize the absolute frequency stabilities of the individual cavities, we first assume that there is no correlation between all three cavities. One can then readily obtain the following equation~\cite{proc1974_Allan}
\begin{equation}
\sigma^2_i(\tau)=[\sigma^2_{ij}(\tau)+\sigma^2_{ik}(\tau)-\sigma^2_{jk}(\tau)]/2.
\label{TCH}
\end{equation}
Here, the subscripts $i$, $j$, and $k$ refer to the three cavities, $\sigma_{ij}(\tau)$ denotes the relative frequency stability between cavity $i$ and $j$, $\sigma_i(\tau)$ denotes the absolute frequency stability of cavity $i$, and $\tau$ is the averaging time, respectively. The absolute frequency stabilities $\sigma_x(\tau)$ ($x=i, j, k$) of the individual cavities are derived from Allan deviations of $\sigma_{ij}(\tau)$, $\sigma_{ik}(\tau)$, and $\sigma_{jk}(\tau)$ using Eq. (\ref{TCH}), and are shown in Fig. \ref{ADSingle}(a). It is clear that cavity $3$ has the best overall frequency stability. It is worth mentioning that even without AVI control, the mid-term stability of cavity $3$ in the $1$-$200$ s region has a mean value of only $2.8\times10^{-15}$, almost two times better than that of cavity $1$ ($5.4\times10^{-15}$), and four times better than that of cavity $2$ ($1.2\times10^{-14}$). This is due to the optimized design of cavity 3. The best frequency stability of $1.3\times10^{-15}$ of cavity 3 is achieved at an averaging time of $0.4$ s.

\subsection{Correlation removed results}
In our experiment, the three ultrastable cavities are located on the same optical table in one sound-proof cabin. There would be unavoidable correlations between them. Neglecting the correlation effects as in section \ref{AllanDev} might contribute errors to the absolute determination of the cavities' performance. We solve this problem by calculating the covariance matrix of the beat signals using methods proposed by Premoli \textit{et al.}~\cite{ieeetim1993_Premoli}. 

For simplicity, we adopt the same notations as used in Ref. \cite{ieeetim1993_Premoli} and summarize the calculations here. Following their approach, cavity 3 is chosen as the reference, and is compared with cavity 1 and cavity 2. The variances and covariances of the two measurement series can be combined into one covariance matrix $S$,
\begin{equation}
S=\left[\matrix{s_{11}&s_{12}\cr
          s_{21}&s_{22}
}\right].
\label{Smatrix}
\end{equation}
Here $s_{11}=\sigma^2_{13}$, $s_{22}=\sigma^2_{23}$, and $s_{12}=s_{21}=\sigma_{13}\sigma_{23}$. From the covariance matrix $S$, and using the relationship
\begin{equation}
\left[\matrix{s_{11}&s_{12}\cr
          s_{21}&s_{22}
}\right]=\left[\matrix{r_{11}+r_{33}-2r_{13}&r_{12}+r_{33}-r_{13}-r_{23}\cr
r_{12}+r_{33}-r_{13}-r_{23}&r_{22}+r_{33}-2r_{23}}\right],
\end{equation}
where $r_{11}$, $r_{22}$, $r_{33}$ are Allan variances of the individual cavities with correlation effect removed, and $r_{12}$, $r_{13}$, $r_{23}$ are correlations between the cavity pairs, we can obtain the following expressions:
\begin{eqnarray}
\label{r11}
r_{11}&=&s_{11}-r_{33}+2r_{13},\nonumber\\ 
r_{12}&=&s_{12}-r_{33}+r_{13}+r_{23},\nonumber\\
r_{22}&=&s_{22}-r_{33}+2r_{23}.
\end{eqnarray}
In order to fix the values of the free parameters $r_{13}$, $r_{23}$, and $r_{33}$, an appropriate criterion ought to be formulated. One possible choice is to require the covariance matrices $S$ and $R$ to be positive definite, where $R$ is defined as 
\begin{equation}
R=\left[\matrix{r_{11}&r_{12}&r_{13}\cr
          r_{12}&r_{22}&r_{23}\cr
          r_{13}&r_{23}&r_{33}
}\right].
\label{Rmatrix}
\end{equation}
Then $r_{13}$, $r_{23}$, and $r_{33}$ must always fulfill the positive definiteness of $R$. This can be done by minimizing the ``global correlation'' among cavities. The detail derivations are rather complicated and can be found in Ref. \cite{ieeetim1993_Premoli}. In the following, we only show the procedure to calculate $r_{13}$, $r_{23}$, and $r_{33}$. We define
\begin{eqnarray}
c_1&=&3\sqrt{\left|S\right|}s_{12}(s_{11}-s_{12})(s_{22}-s_{12}),\nonumber\\ 
c_2&=&2.25\left|S\right|^2+2(s_{11}+s_{22}+s_{12})c_1/(3\sqrt{\left|S\right|}),\nonumber\\
c_3&=&3\left|S\right|^{3/2}(s_{11}+s_{22})+c_1/3,\nonumber\\
c_4&=&\left|S\right|[1.5\left|S\right|+(s_{11}+s_{22}-s_{12})(s_{11}+s_{22}+s_{12})],\nonumber\\
c_5&=&\left|S\right|^{3/2}(s_{11}+s_{22}),\nonumber\\
c_6&=&\left|S\right|^2/4,
\end{eqnarray}
and solve 
\begin{equation}
c_1f+c_2f^2+c_3f^3+c_4f^4+c_5f^5+c_6f^6=0
\end{equation}
for the unknown $f$. We then choose the unique minimum positive root $f_{min}$ if it exists; otherwise, choose the $f_{min}=0$. With $f=f_{min}$, we can calculate 
\begin{eqnarray}
b_0&=&\sqrt{\left|S\right|}s^2_{12}+[s^2_{12}(s_{11}+s_{22})]f+\sqrt{\left|S\right|}s^2_{12}f^2,\nonumber\\ 
b_1&=&-\sqrt{\left|S\right|}s_{12}-(2s^2_{12}+3\left|S\right|/2)f-\sqrt{\left|S\right|}(s_{11}+s_{22})f^2-\left|S\right|f^3/2,\nonumber\\
b_2&=&\sqrt{\left|S\right|}+2(s_{11}+s_{22}-s_{12})f+3\sqrt{\left|S\right|}f^2.
\end{eqnarray}
The free parameters $r_{13}$, $r_{23}$, $r_{33}$ are determined to be 
\begin{eqnarray}
r_{33}&=&-b_1/b_2,\nonumber\\
r_{13}&=&r_{33}-\frac{a_{10}(a_{02}-a_{11})}{a_{20}a_{02}-a^2_{11}},\nonumber\\
r_{23}&=&r_{33}-\frac{a_{10}(a_{20}-a_{11})}{a_{20}a_{02}-a^2_{11}},
\end{eqnarray}
where
\begin{eqnarray}
a_{20}&=&2\sqrt{\left|S\right|}+fs_{22},\nonumber\\
a_{02}&=&2\sqrt{\left|S\right|}+fs_{11},\nonumber\\
a_{11}&=&\sqrt{\left|S\right|}-fs_{12},\nonumber\\
a_{10}&=&a_{01}=\sqrt{\left|S\right|}(2r_{33}+s_{12}).
\end{eqnarray}
Finally, $r_{11}$, $r_{12}$, and $r_{22}$ can be calculated from Eq. (\ref{r11}). The Allan deviation of a cavity $x$ ($x=1, 2, 3$) can then be calculated by using the diagonal matrix elements of $R$, 
\begin{equation}
\sigma_x=\sqrt{r_{xx}}.
\end{equation}

\begin{figure}[htb]
\centering\includegraphics[width=7cm]{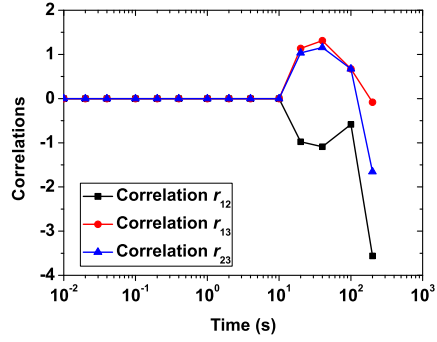}
\caption{Estimated correlations for the three reference cavity pairs.}
\label{Covariance}
\end{figure}

The results are shown in Fig.~\ref{ADSingle}(b). By comparing Fig.~\ref{ADSingle}(a) and (b), we can see that the Allan deviations of the three reference cavities with correlations removed are slightly better than those that we obtained from the classical three-cornered-hat method. The frequency stability for cavity $3$ is $1.4\times10^{-15}$ at $1$ s. The correlations of the three cavity pairs are also shown in Fig.~\ref{Covariance}. At time scale shorter than $10$ s, the $R$ matrix is diagonal with zero estimated correlations between cavities. Correlations appear after $10$ s, most probably due to common temperature fluctuations and other environmental changes in lab 1. 

It should be emphasized here that choosing cavity 3 as the reference is totally arbitrary. We have also chosen cavity 1 or cavity 2 as the reference, and repeated the above calculations. The results are the same.

\subsection{Gaussian process criteria}
It is interesting to check whether the beat signals from the three reference cavities obey Gaussian distribution. This can be examined through the Gaussian moment theorem, which states that all higher order correlations among Gaussian variables are expressible in terms of second-order correlations between all pairs of variables. For example, the fourth-order correlation between four such variables can be expressed as
\begin{eqnarray}
<\Delta x_1\Delta x_2\Delta x_3\Delta x_4>&=&<\Delta x_1\Delta x_2><\Delta x_3\Delta x_4>+<\Delta x_1\Delta x_3><\Delta x_2\Delta x_4> \nonumber \\
&&+<\Delta x_1\Delta x_4><\Delta x_2\Delta x_3>.
\end{eqnarray}
Hence, if we assume that the beat signals were Gaussian processes, then we should have
\begin{equation}
\xi^4_{ij}\equiv<(y_i-y_j)^4>=3<(y_i-y_j)^2>^2=3(\sigma^2_{ij})^2,
\end{equation}
so that the fourth-order correlation $\xi^4_{ij}$ of any beat signals is equal to three times the respective Allan variance squared.

\begin{figure}[htb]
\centering\includegraphics[width=7cm]{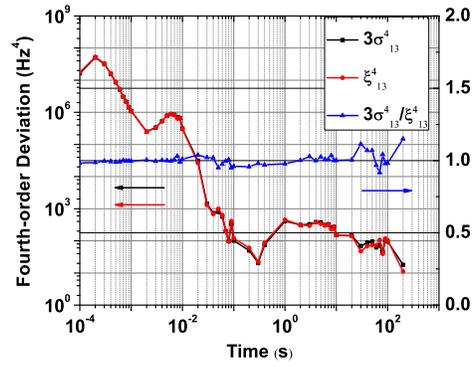}
\caption{Comparison between the fourth-order correlation and three times the Allan variance squared of one set of the three beat signals (between cavity 1 and 3) confirms the Gaussian nature of the signals.}
\label{Gaussian}
\end{figure}

Fig.~\ref{Gaussian} compares the fourth-order correlation and three times the Allan variance squared, for one set of such three beat signals (between cavity 1 and 3). The comparison shows that although there is a slight difference between the two data sets, the ratio of them remains very close to 1 during the entire time scale. We may also notice that the difference becomes more obvious at longer time scales ($>10$ s), which mainly comes from the increasing uncertainty caused by fewer data points at longer times. This result confirms the fact that beat signals are composed of a multitude of independent additive contributions, which as a result of the central limit theorem leads to Gaussian statistics.

\section{Conclusion}
In conclusion, we have measured the absolute frequency stabilities of three laser beams independently locked to three ultrastable reference cavities, by using both a classical three-cornered-hat method and a modified three-cornered-hat method, where two primary light sources are used to avoid common phase noise. One of the reference cavities is designed to be vibration insensitive that does not even require vibration isolation under normal laboratory seismic conditions. Our measurements clearly demonstrate that this cavity has the best frequency stability over the entire measurement time from $100$ $\mu$s to $200$ s. A frequency stability of $1.3\times10^{-15}$ at $0.4$ s is observed for this cavity, even without vibration isolation. We further investigate correlations between the reference cavities at different time scales, and observe that correlations between cavities start to appear from $10$ s to $200$ s according to calculated covariance matrix. The correlations are mainly caused by the same laboratory surroundings of the three cavities. An absolute, correlation-removed Allan deviation of $1.4\times10^{-15}$ at $1$ s of the newly designed cavity is obtained. This gives a frequency uncertainty of only $0.44$ Hz.  

\section*{Acknowledgments}
We thank J. Stuhler, G. Schuricht, and F. Lison of Toptica Photonics for helpful discussions, and R. Lalezari of Advance Thin Films for technical suggestions. Z. H. Lu acknowledges fellowship support from the Alexander von Humboldt Foundation.

\end{document}